\documentclass[journal]{IEEEtran}

\usepackage{eso-pic}
\usepackage[utf8]{inputenc}
\usepackage[T1]{fontenc}
\usepackage{amsmath,amssymb}
\usepackage{bm}
\usepackage{graphicx}
\usepackage{subcaption}
\usepackage{booktabs}
\usepackage{cite}
\usepackage[hidelinks]{hyperref}
\usepackage{xcolor}
\begin{document}
	\title{Interpretable Feature Learning for RF Fingerprinting via Polar MKANs}
	\author{\IEEEauthorblockN{%
			Mikhail Krasnov$^{\dag}$, Ljupcho Milosheski$^{\dag}$, Carolina Fortuna \\
			Jo\v{z}ef Stefan Institute, Ljubljana, Slovenia \\
			\{mikhail.krasnov, ljupcho.milosheski, carolina.fortuna\}@ijs.si
		}\\
		\thanks{$^{\dag}$Mikhail Krasnov and Ljupcho Milosheski contributed equally to this work.}%
	}
	\markboth{IEEE Wireless Communications Letters}{}
	\maketitle
	\AddToShipoutPictureFG*{%
		\AtPageUpperLeft{%
			\put(0,-18){%
				\makebox[\paperwidth][c]{%
					\footnotesize\itshape
					This work has been submitted to the IEEE for possible publication.
					Copyright may be transferred without notice, after which this version
					may no longer be accessible.
				}%
			}%
		}%
	}
	\begin{abstract}
		Radio frequency (RF) fingerprinting authenticates wireless devices from hardware-induced I/Q impairments, typically with deep-learning feature extractors that are accurate but opaque, limiting their use in security-critical settings. We propose Polar Monotonic Kolmogorov--Arnold Networks (Polar MKAN), a block-partitioned monotonic encoder on polar inputs in which each latent dimension depends exclusively on magnitude or phase, yielding channel separation and monotone responses by construction. On a synthetic gain/carrier frequency offset (CFO) benchmark, Polar MKAN reaches $57.2\%$ DCI Disentanglement versus $\leq12.9\%$ for unpartitioned baselines. We further evaluate the detection-accuracy trade-off on real data and the sensitivity to blind CFO compensation.
	\end{abstract}
	\section{Introduction}
	\label{sec:intro}
	RF fingerprinting exploits hardware-level imperfections such as carrier frequency offset (CFO), I/Q imbalance, and power-amplifier nonlinearities~\cite{sankhe2019no}, to authenticate wireless devices without cryptographic overhead~\cite{soltanieh2020review,xie2024radio}. Unknown emitter detection (UED) extends this capability to open-set scenarios in which previously unseen devices must be identified at inference time~\cite{robinson2021novel,apfeld2021recognition}. Self-supervised approaches based on DL architectures~\cite{zhan2023mcrff,hao2023contrastive} are particularly suitable for this task due to their ability to learn from large amounts of unlabelled data, and recent work shows strong UED performance from such self-supervised feature spaces~\cite{krasnov2025design}. However, UED systems based on standard DL architectures remain opaque and intractable for an operator due to the large number of parameters standing between the latent features and any interpretable signal property. For security-sensitive applications such as spectrum monitoring and critical-infrastructure protection, this lack of transparency is a deployment barrier~\cite{rudin2022interpretable}.
	Existing approaches to UED interpretability rely on post-hoc methods such as gradient saliency~\cite{simonyan2013deep}, LIME~\cite{ribeiro2016should}, and SHAP~\cite{lundberg2017unified}, which provide local, sample-specific explanations without structural guarantees about global consistency. A complementary thread pursues \emph{disentangled representation learning} for RF fingerprinting~\cite{xie2023disentangled,zhang2025factorized,pan2025cross}, separating device-relevant information from channel, receiver, modulation, and other nuisance factors through adversarial or regularised objectives. These methods share a fundamental limitation, namely that disentanglement is \emph{encouraged} but not guaranteed in general~\cite{locatello2019challenging}, and the separation quality depends on training dynamics that cannot be verified at inference time.
	Kolmogorov--Arnold networks (KANs)~\cite{liu2024kan,somvanshi2024survey} offer an architectural route, since their visible spline functions enable global interpretability~\cite{krasnov2025design}, but non-monotonic activations still yield complex input/output mappings. MKANs~\cite{krasnov2026monotonic} enforce hard monotonicity constraints by making each spline activation monotonic, which fixes the sign of every input's effect on every output and so makes the input/output mapping predictable. Certified-monotonic alternatives such as MonoKAN~\cite{polo2025monokan} exist, but we build on MKAN since its representation-learning guarantee (Theorem~1 of~\cite{krasnov2026monotonic}) is established in the deep-clustering setting, certifying that a monotone extractor preserves the neighbourhood structure the clustering-based UED decision relies on.
	To the best of our knowledge, no prior work achieves an \emph{architecturally enforced} separation of the magnitude and phase information channels in the RF-fingerprinting feature space. Building on the MKAN construction and UED workflow of~\cite{krasnov2026monotonic,krasnov2025design}, this letter's new contributions are the polar input representation and the block-partitioned output, summarized as follows.
	\begin{itemize}
		\item An encoder whose latent space is split by construction, with dedicated latent dimensions wired exclusively to the magnitude and to the principal-phase inputs, so each feature reflects exactly one of the two physical channels.
		\item A synthetic-benchmark evaluation with known ground-truth device impairments, on which Polar MKAN recovers the magnitude/phase separation ($57.2\%$ DCI Disentanglement~\cite{eastwood2018framework,carbonneau2022measuring} vs.\ $\le 12.9\%$ for unconstrained and polar-input-only baselines), and a quantified interpretability-performance trade-off on the UED task.
		\item A perturbation analysis on real WiSig\cite{hanna_wisig_2022} signals showing that the phase channels respond monotonically and stay separated from the magnitude channel, so a feature-space deviation decomposes into a magnitude- and a phase-channel component.
	\end{itemize}
	\section{Background and System Model}
	\label{sec:background}
\begin{figure}[t]
		\centering
		\includegraphics[width=0.9\columnwidth]{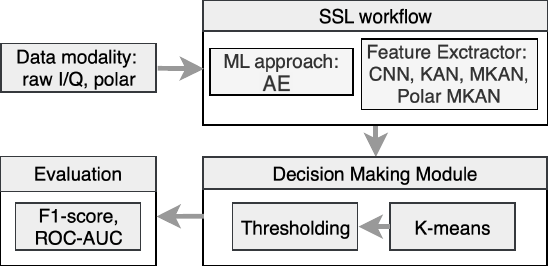}
		\caption{Self-supervised UED workflow.}
		\label{fig:workflow}
	\end{figure}

	We adopt the workflow of~\cite{krasnov2025design} depicted also in Fig.~\ref{fig:workflow}. A data modality (raw I/Q or polar coordinates) is mapped by a feature extractor (FE) $F:\mathbb{R}^{2N}\rightarrow\mathbb{R}^{N^*}$ to a low-dimensional feature space. The FE is trained as the encoder of a self-supervised autoencoder, and at inference a cluster-based decision module compares test features with training clusters for UED. We keep the training and decision procedure fixed and modify only the FE and its input representation.
	\subsection{Monotonic KAN}
	\label{sec:mkan}
	A KAN layer~\cite{liu2024kan} replaces fixed multi-layer perceptron (MLP) activations with learnable B-spline functions $\phi_{ij}(x_i)$ on each edge, giving
	\begin{equation}
		F^{(j)}(\mathbf{x})
		=
		\sum_i w^{(s)}_{ij}\phi_{ij}(x_i)
		+
		w^{(b)}_{ij}\operatorname{SiLU}(x_i).
	\end{equation}
	MKANs~\cite{krasnov2026monotonic} constrain spline coefficients to be ordered, reparameterize connected weights through exponentiation, and replace the non-monotone SiLU base with ReLU, giving
	\begin{equation}
		\widetilde F^{(j)}(\mathbf{x})
		=
		\sum_i e^{\widetilde w^{(s)}_{ij}}\widetilde\phi_{ij}(x_i)
		+
		e^{\widetilde w^{(b)}_{ij}}\operatorname{ReLU}(x_i)+b_j,
	\end{equation}
	where $\widetilde\phi_{ij}$ are monotone splines. Hence an output is componentwise nondecreasing in every input to which it is connected, while disconnected inputs have zero effect. Theorem~1 of~\cite{krasnov2026monotonic} shows that an increasing FE can reproduce the semantic-neighbourhood partition of an arbitrary FE with at most $2\times$ feature dimensionality.
	\subsection{Proposed Polar MKAN}
	\label{sec:polar_mkan}
\begin{figure}[t]
		\centering
		\begin{subfigure}[b]{0.4\columnwidth}
			\includegraphics[width=\textwidth]{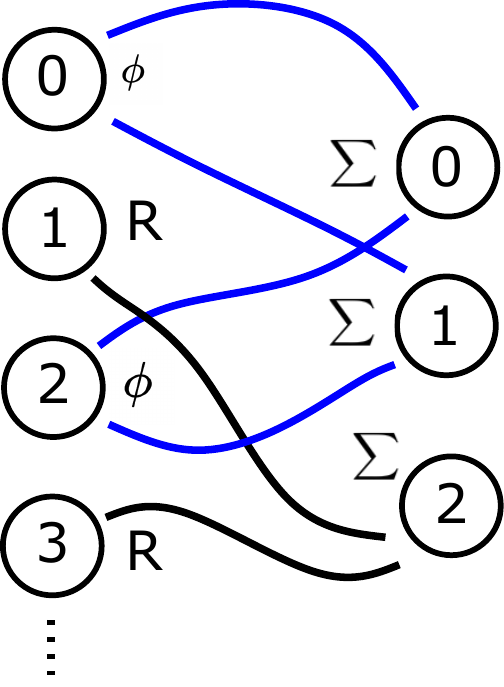}
			\caption{Architecture}
			\label{fig:polar_arch}
		\end{subfigure}
		\hfill
		\begin{subfigure}[b]{0.5\columnwidth}
			\includegraphics[width=\textwidth]{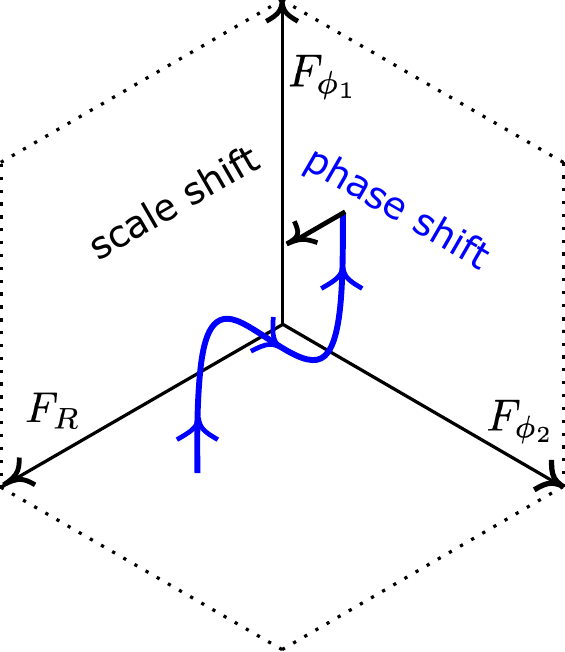}
			\caption{Feature-space geometry}
			\label{fig:polar_fs}
		\end{subfigure}
		\caption{Polar MKAN. (a)~Exclusive connections, where $F_{\phi_1},F_{\phi_2}$ receive only phase inputs and $F_R$ only magnitude inputs. (b)~Idealised feature-space geometry away from the phase branch cut.}
		\label{fig:polar_architecture}
	\end{figure}

	For $x^I[i]+jx^Q[i]$, we form principal-polar coordinates
	\begin{equation}
		x^I[i]+jx^Q[i]=R[i]e^{j\phi[i]},
		\qquad
		\phi[i]\in[0,2\pi).
	\end{equation}
	The implementation feeds polar models the fixed numerical scaling
	$\widetilde R[i]=R[i]/1.5$ and $\widetilde\phi[i]=\phi[i]/(2\pi)$.
	Only Polar MKAN imposes exclusive output connectivity, as illustrated in Fig.~\ref{fig:polar_arch}, where $F_R$ receives the magnitude block $\{\widetilde R[i]\}$ while $F_{\phi_1}$ and $F_{\phi_2}$ receive the phase block $\{\widetilde\phi[i]\}$. Two outputs are allocated to phase because a substantial, often dominant, share of the discriminative fingerprint information is phase-encoded~\cite{soltanieh2020review}. Polar CNN and Polar KAN use the same polar input, but without partitioned connectivity. The Polar MKAN architecture thus enforces channel-level separation without assuming that the two phase outputs isolate distinct physical factors.
	Formally,
	\begin{equation}
		\label{eq:split}
		\begin{split}
			F_{\phi_1}&=F_{\phi_1}(\widetilde\phi_1,\ldots,\widetilde\phi_N),\\
			F_{\phi_2}&=F_{\phi_2}(\widetilde\phi_1,\ldots,\widetilde\phi_N),\\
			F_R&=F_R(\widetilde R_1,\ldots,\widetilde R_N).
		\end{split}
	\end{equation}
	Using $\succeq$ for componentwise ordering, monotonicity and the exclusive connections give the forward guarantees
	\begin{equation}
		\label{eq:monotone_order}
		\begin{split}
			\widetilde{\bm\phi}'\succeq\widetilde{\bm\phi},\
			\widetilde{\mathbf R}'=\widetilde{\mathbf R}
			&\Rightarrow
			F'_{\phi_k}\geq F_{\phi_k},\ k\in\{1,2\},\quad F'_R=F_R,\\
			\widetilde{\mathbf R}'\succeq\widetilde{\mathbf R},\
			\widetilde{\bm\phi}'=\widetilde{\bm\phi}
			&\Rightarrow
			F'_R\geq F_R,\quad F'_{\phi_k}=F_{\phi_k}.
		\end{split}
	\end{equation}

	This means that raising every phase input while holding the magnitudes fixed can only raise the phase features, and leaves $F_R$ untouched since it has no connection to the phase inputs. The second line mirrors this for magnitude. Monotonicity thus fixes the direction a feature may move, and the exclusive connections fix which features may move at all.
	Two conditions delimit these guarantees: (i) they run from an ordered input change to a feature change, so $\Delta F_\phi>0$ indicates a rise of the phase inputs, irrespective of its physical origin; (ii) they are stated in the principal-phase representation, so a phase shift is covered as long as every sample's phase stays below $2\pi$, since a sample that wraps back to $0$ has decreased in this representation and the premise $\widetilde{\bm\phi}'\succeq\widetilde{\bm\phi}$ then fails.
	\section{Evaluation on Simulated/Synthetic Dataset}
	\label{sec:synth}
	We construct a controlled RF-fingerprinting benchmark in which devices differ through fixed impairment parameters CFO, gain, I/Q imbalance, and AWGN model. No existing measured RF dataset has annotations that enables the study of disentanglement.
	We use the same impairment model for two complementary synthetic dataset evaluations: feature--factor analysis and UED evaluation. The feature--factor analysis uses 1,000 training and 100 held-out evaluation devices with one burst each (Table~\ref{tab:dataset}), which gives broad coverage of the impairment space. The synthetic UED experiment uses 200 devices with 20 bursts each, since training, clustering, and open-set testing require repeated observations per device.
	\begin{table}[t]
		\centering
		\caption{Composition of the synthetic dataset enabling a controlled disentanglement study.}
		\label{tab:dataset}
		\begin{tabular}{lcc}
			\toprule
			& \textbf{Train} & \textbf{Evaluation} \\
			\midrule
			Devices & $1000$ & $100$ \\
			Bursts/device & $1$ & $1$ \\
			Burst length ($N$) & $256$ & $256$ \\
			Samples/symbol ($L$) & $4$ & $4$ \\
			SNR range (dB) & $[10,30]$ & $[10,30]$ \\
			\bottomrule
		\end{tabular}
	\end{table}
	\paragraph{Signal and impairment model}
	Every burst uses the same fixed unit-power 16-QAM symbol sequence with levels $\{-3,-1,1,3\}$ on both axes, upsampled by $L=4$ and root-raised-cosine pulse-shaped (roll-off $0.35$, span 8). The fixed sequence removes payload variation.
	Each device $d$ has CFO $\nu_d\sim\mathcal N(0,(2\times10^{-3})^2)$ and gain perturbation $g_d\sim\mathcal N(0,0.07^2)$. With pulse-shaped $s[n]$,
	\begin{equation}
		u_d[n]=s[n]e^{j2\pi\nu_dn}.
	\end{equation}
	We additionally introduce complex I/Q imbalance as structured nuisance variation. Let $a_d\sim\mathcal N(0,0.3^2)$ dB, $\varphi_d\sim\mathcal N(0,0.03^2)$ rad, and $\gamma_d=10^{a_d/20}$. The equivalent-baseband coefficients are
	\begin{equation}
		\mu_d=\tfrac12(1+\gamma_de^{-j\varphi_d}),
		\qquad
		\eta_d=\tfrac12(1-\gamma_de^{j\varphi_d}),
	\end{equation}
	giving
	\begin{equation}
		x_d[n]=(1+g_d)\left[\mu_du_d[n]+\eta_du_d^*[n]\right].
		\label{eq:impairments_iq}
	\end{equation}
	At zero imbalance ($\gamma_d=1,\varphi_d=0$), $\mu_d=1$ and $\eta_d=0$. The imbalance is applied after the CFO/gain operation, matching the implementation. Because the image term contains $u_d^*[n]$, nonzero I/Q imbalance couples magnitude and phase, so the CFO effect is no longer purely phase-only.
	
	Finally,
	\begin{equation}
		y_d[n]=x_d[n]+w[n],
		\qquad
		w[n]\sim\mathcal{CN}(0,\sigma_w^2),
	\end{equation}
	where $\sigma_w^2=P_x/10^{\mathrm{SNR}/10}$ and $\mathbb E|w[n]|^2=\sigma_w^2$. Raw-I/Q models receive $(I,Q)$ directly, and polar models convert each burst to $(\widetilde R,\widetilde\phi)$ at loading time. The scored factor vector is $\bm\theta_d=(g_d,\nu_d)$, and the I/Q-imbalance parameters act as unscored structured nuisance variables.
	
	\begin{table}[t]
		\centering
		\caption{DCI Disentanglement (D) and Completeness (C) (\%, mean $\pm$ 95\% CI over 10 runs).}
		\label{tab:architectures_disent}
		\begin{tabular}{lcc}
			\toprule
			Architecture & D & C \\
			\midrule
			CNN        & $6.1\pm4.2$   & $6.7\pm3.4$ \\
			KAN        & $10.0\pm4.7$  & $9.8\pm4.0$ \\
			MKAN       & $7.4\pm3.1$   & $9.5\pm2.9$ \\
			Polar CNN  & $5.0\pm2.4$   & $5.3\pm1.8$ \\
			Polar KAN  & $12.9\pm4.8$  & $13.1\pm4.9$ \\
			Polar MKAN & $\mathbf{57.2\pm6.0}$ & $\mathbf{47.0\pm4.0}$ \\
			\bottomrule
		\end{tabular}
	\end{table}
	\paragraph{Feature--factor analysis}
	All six autoencoders\footnote{Code and configurations:
		\url{https://github.com/sensorlab/PolarMKAN}} are trained for 120 epochs with Adam (learning rate $10^{-2}$), $N^*=3$, on the same synthetic training data. Polar CNN and Polar KAN serve as polar-input-only controls with unpartitioned encoders, while only the design of Polar MKAN enables the exclusive magnitude/phase output structure from Figure \ref{fig:polar_architecture}.
	
	Following the DCI framework~\cite{eastwood2018framework,carbonneau2022measuring}, the evaluation estimates an $N^*\times K$ feature-importance matrix $R$ by fitting one gradient-boosted regressor (100 estimators, maximum depth 3) from the learned coordinates to each of the $K=2$ scored factors. For latent coordinate $i$,
	\begin{equation}
		D_i=1-\frac{H(P_{i\cdot})}{\log K},
		\qquad
		P_{ij}=\frac{R_{ij}}{\sum_{k}R_{ik}},
	\end{equation}
	and overall disentanglement uses the canonical importance weighting
	\begin{equation}
		D=\sum_i\rho_iD_i,
		\qquad
		\rho_i=\frac{\sum_jR_{ij}}{\sum_{k,j}R_{kj}}.
	\end{equation}
	Completeness (C) is computed analogously across latent coordinates for each factor. Without I/Q imbalance, gain and CFO would align exactly with magnitude and phase, respectively. Under the full model used here, the image term introduces cross-channel coupling, so the correspondence is no longer exact. The experiment therefore evaluates whether the imposed magnitude/phase structure remains informative about the two device factors in the presence of structured nuisance variation.
	Table~\ref{tab:architectures_disent} shows that only Polar MKAN achieves substantial $57.2\%$ D and $47.0\%$ C values, while Polar KAN and Polar CNN show that polar coordinates alone do not recover comparable separation, achieving $12.9\%$, $5\%$ D and $13.1\%$, $5.3\%$, respectively. The raw-I/Q MKAN shows that monotonicity in Cartesian coordinates is insufficient. The lower completeness, as shown in the last column of \ref{tab:architectures_disent} is consistent with the two phase coordinates that share information about the single scored CFO factor.
	\paragraph{Phase response validation}
	
\begin{figure}[t]
		\centering
		\includegraphics[width=0.7\columnwidth]{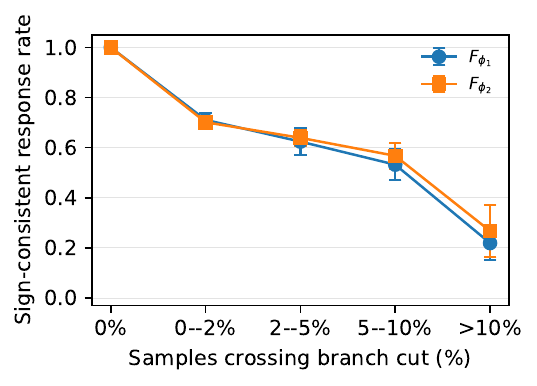}
		\caption{Sign consistency of the phase features vs.\ fraction of samples whose phase wraps (95\% CI, 10 runs).}
		\label{fig:phase_wrap}
	\end{figure}

	Eq.~\eqref{eq:monotone_order} guarantees the response sign only for wrap-free inputs, so the practically relevant property is how quickly sign consistency is lost once samples start crossing the $2\pi$ branch cut. To characterise this, we apply rotations in $[-0.6,0.6]$ rad to the synthetic evaluation set over 10 independently trained Polar MKANs and group the responses by the fraction of samples whose phase wraps, with the results shown in Fig.~\ref{fig:phase_wrap}. Sign consistency decreases gradually with the wrapped fraction, from complete consistency when no sample wraps to roughly $70\%$ below $2\%$ wrapped samples, and to $20$--$30\%$ once more than $10\%$ of the samples cross the $2\pi$. The sign reading of the phase features therefore remains reliable for small rotations, and degrades gradually as wrapping becomes widespread.
	\paragraph{Synthetic UED}
	Factor alignment alone does not establish that the representation remains useful for detection, so we also run the clustering-based UED task of Fig.~\ref{fig:workflow} on the synthetic data, mirroring the WiSig protocol of Section~\ref{sec:interpretability}. A fixed 25\% of the 200 devices are unknown and appear only in the test set, and for the rest 80\% of bursts train and 20\% test. Models train for 120 epochs with Adam (learning rate $10^{-2}$), batch size 50 and input-noise standard deviation 0.01, using 80 clusters. Results are averaged over 10 training seeds.
	\paragraph{Performance and interpretability}
	\begin{figure}[t]
		\centering
		\includegraphics[width=0.82\columnwidth]{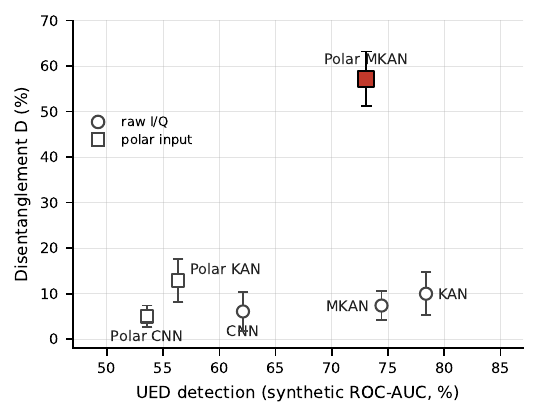}
		\caption{Synthetic UED ROC AUC vs.\ held-out DCI Disentanglement (95\% CI, 10 runs).}
		\label{fig:tradeoff}
	\end{figure}
	Fig.~\ref{fig:tradeoff} compares synthetic UED performance with held-out DCI Disentanglement. Polar MKAN occupies a distinct high-disentanglement region while retaining $73.1\%$ synthetic UED ROC AUC, within one standard deviation of raw-I/Q MKAN at $74.4\%$ (Table~\ref{tab:ued}). Polar KAN, with the same input but neither monotonicity nor exclusive connections, reaches only $12.9\%$ D, and raw-I/Q MKAN $7.4\%$, so the separation is associated with the \emph{combined} block-partitioned monotonic construction rather than the polar transform alone.
	Taken together, the synthetic experiments evaluate two complementary properties. DCI measures alignment with known device factors, while UED measures whether the representation preserves device separability. Polar MKAN substantially improves the former while retaining useful detection performance.
	\section{Validation and Evaluation on WiSig}
	\label{sec:interpretability}
	We validate and evaluate the proposed method versus selected alternatives on WiSig~\cite{hanna_wisig_2022} SingleDay and ManyTx subsets, using length-256 I/Q and a single receiver (index 0) in both. SingleDay contains 28 emitters from one capture day, and ManyTx contains 150 emitters and samples from four days. Device-level open-set folds are generated by the same repository routine used in~\cite{krasnov2025design} (fold-ratio settings 7 and 5, respectively), and devices returned in each fold's unknown list are excluded from that fold's training data and labelled unknown at evaluation. Models are trained for 120 epochs with Adam (learning rate $10^{-3}$), batch size 50, using 280 clusters for SingleDay and 1,000 for ManyTx. The number of clusters is set to $5$--$10\times$ the number of classes, following the findings in \cite{krasnov2025novel}. Each fold is trained once, so the reported WiSig std aggregates device-split and training variability. 
	
	\paragraph{Amplitude/phase response}
	\begin{figure}[t]
		\centering
		\begin{subfigure}[b]{0.8\columnwidth}
			\includegraphics[width=0.9\textwidth]{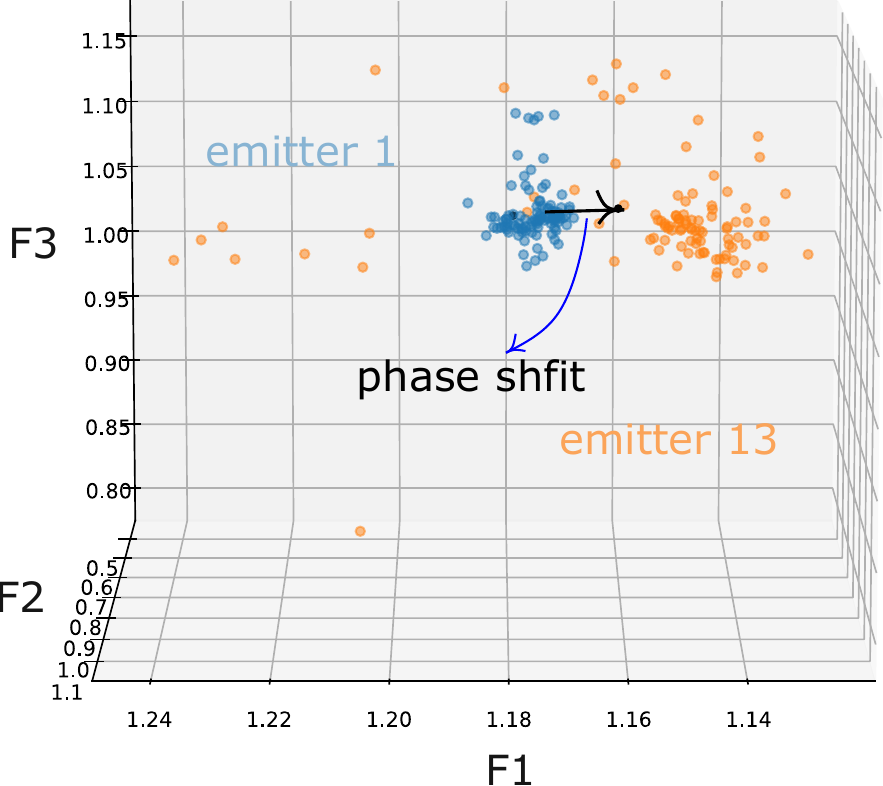}
			\caption{Raw I/Q MKAN}
			\label{fig:mkan_phase}
		\end{subfigure}\\
		\begin{subfigure}[b]{0.98\columnwidth}
			\centering
			\includegraphics[width=0.9\textwidth]{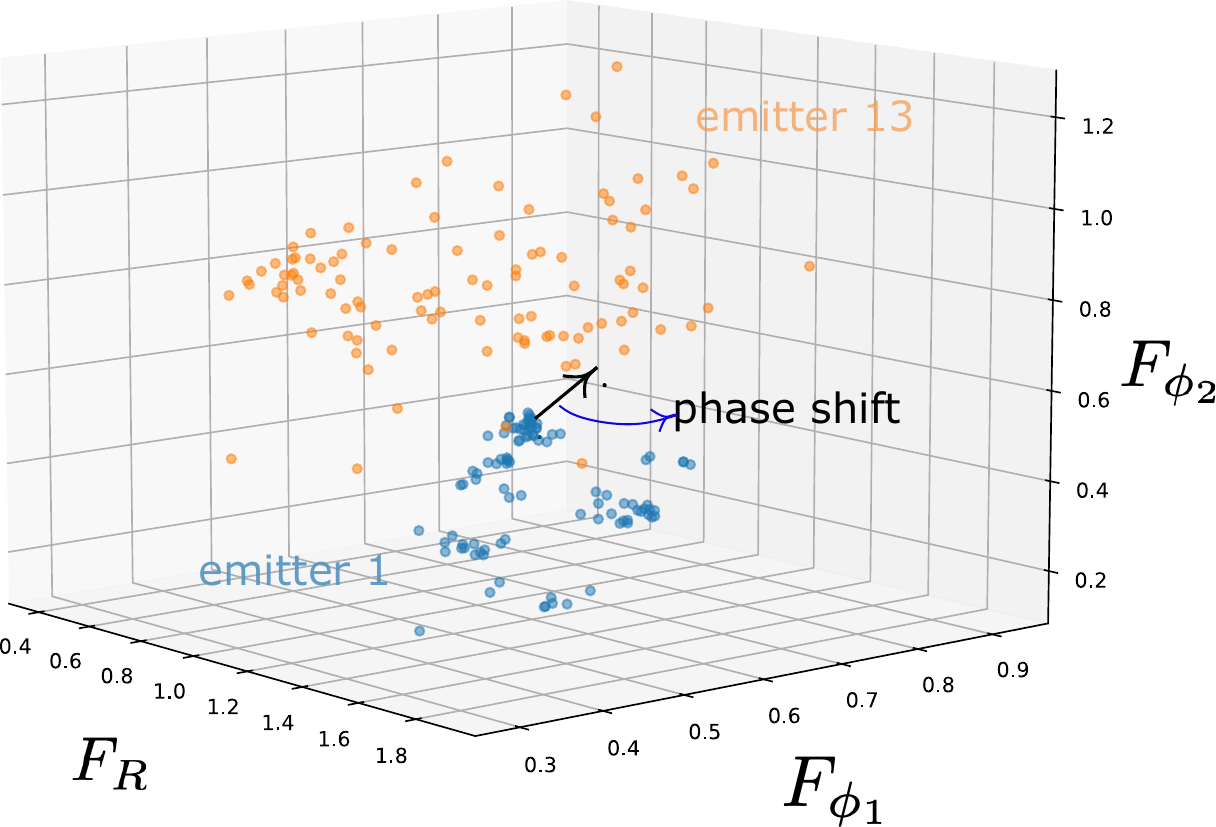}
			\caption{Polar MKAN}
			\label{fig:polar_mkan_phase}
		\end{subfigure}
		\caption{Response to a positive carrier-phase rotation on WiSig ManyTx.}
		\label{fig:phase_shift}
	\end{figure}
	Fig.~\ref{fig:phase_shift} illustrates the response on real signals. A common positive rotation leaves magnitudes unchanged. In raw I/Q it does not produce componentwise ordered inputs, so an MKAN can move latent coordinates in different directions, whereas in Polar MKAN a wrap-free rotation obeys~\eqref{eq:monotone_order}, so $F_R$ is unchanged and both phase outputs are nondecreasing.
	For a flagged sample, its deviation from the nearest known cluster centre, $(\Delta F_R,\Delta F_{\phi_1},\Delta F_{\phi_2})$, therefore decomposes the anomaly into a magnitude-channel and a phase-channel component, measured in units of within-cluster spread, and a positive $\Delta F_{\phi}$ indicates an aggregate positive phase rotation relative to the cluster, with the sign interpretation inherited from~\eqref{eq:monotone_order} under its wrap-free condition. In an unconstrained feature space, the same deviation vector points in an arbitrary learned direction and supports no such reading.
	\begin{table*}[t]
		\centering
		\caption{UED results (\%). AUC and F1 uncompensated, AUC$_{\mathrm{comp}}$ with blind CFO compensation before training and evaluation. Std over 10 seeds (synthetic) or device folds (WiSig). Best uncompensated in bold.}
		\label{tab:ued}
		\setlength{\tabcolsep}{4pt}
		\begin{tabular}{@{}l ccc ccc ccc@{}}
			\toprule
			& \multicolumn{3}{c}{Synthetic} & \multicolumn{3}{c}{SingleDay} & \multicolumn{3}{c}{ManyTx} \\
			\cmidrule(lr){2-4}\cmidrule(lr){5-7}\cmidrule(lr){8-10}
			FE & AUC & F1 & AUC$_{\mathrm{comp}}$ & AUC & F1 & AUC$_{\mathrm{comp}}$ & AUC & F1 & AUC$_{\mathrm{comp}}$ \\
			\midrule
			CNN        & 62.1{\scriptsize$\pm$2.2} & 24.8{\scriptsize$\pm$3.6} & 51.1{\scriptsize$\pm$1.7} & 71.0{\scriptsize$\pm$2.5} & 36.8{\scriptsize$\pm$5.4} & 66.8{\scriptsize$\pm$2.4} & 64.7{\scriptsize$\pm$1.8} & 28.8{\scriptsize$\pm$3.9} & 62.3{\scriptsize$\pm$1.3} \\
			KAN        & \textbf{78.4}{\scriptsize$\pm$4.2} & \textbf{62.9}{\scriptsize$\pm$8.7} & 51.1{\scriptsize$\pm$1.1} & \textbf{89.8}{\scriptsize$\pm$3.6} & \textbf{74.0}{\scriptsize$\pm$7.0} & 76.2{\scriptsize$\pm$4.6} & \textbf{73.2}{\scriptsize$\pm$2.6} & \textbf{44.0}{\scriptsize$\pm$5.5} & 67.1{\scriptsize$\pm$3.1} \\
			MKAN       & 74.4{\scriptsize$\pm$4.6} & 53.3{\scriptsize$\pm$8.2} & 49.8{\scriptsize$\pm$1.4} & 79.7{\scriptsize$\pm$5.8} & 61.3{\scriptsize$\pm$7.9} & 66.5{\scriptsize$\pm$5.1} & 69.9{\scriptsize$\pm$1.2} & 38.2{\scriptsize$\pm$2.4} & 60.8{\scriptsize$\pm$2.3} \\
			Polar CNN  & 53.6{\scriptsize$\pm$1.6} & 20.6{\scriptsize$\pm$2.2} & 50.4{\scriptsize$\pm$0.9} & 64.8{\scriptsize$\pm$2.9} & 27.4{\scriptsize$\pm$5.9} & 62.6{\scriptsize$\pm$2.9} & 61.7{\scriptsize$\pm$1.1} & 24.1{\scriptsize$\pm$2.0} & 60.9{\scriptsize$\pm$1.0} \\
			Polar KAN  & 56.3{\scriptsize$\pm$5.3} & 23.2{\scriptsize$\pm$7.5} & 50.5{\scriptsize$\pm$1.7} & 82.8{\scriptsize$\pm$4.6} & 59.8{\scriptsize$\pm$7.1} & 71.0{\scriptsize$\pm$5.8} & 66.3{\scriptsize$\pm$1.8} & 30.9{\scriptsize$\pm$4.0} & 62.2{\scriptsize$\pm$2.4} \\
			Polar MKAN & 73.1{\scriptsize$\pm$3.6} & 45.9{\scriptsize$\pm$4.6} & 49.8{\scriptsize$\pm$1.1} & 78.8{\scriptsize$\pm$4.8} & 52.6{\scriptsize$\pm$9.1} & 66.5{\scriptsize$\pm$4.6} & 67.2{\scriptsize$\pm$1.0} & 31.5{\scriptsize$\pm$2.1} & 61.3{\scriptsize$\pm$1.9} \\
			\bottomrule
		\end{tabular}
	\end{table*}
	\paragraph{Performance--interpretability trade-off}
	Polar MKAN reaches $78.8\%$ ROC AUC on SingleDay (KAN $89.8\%$) and $67.2\%$ on ManyTx (KAN $73.2\%$) as summarized in Table \ref{tab:ued}, quantifying the real-data cost of the stronger structural bias. The constrained model therefore trades approximately 11 AUC points on SingleDay and 6 points on ManyTx for the explicit magnitude/phase feature structure. Part of this gap is a dimensionality effect, since by Theorem~1 of~\cite{krasnov2026monotonic}, a monotone FE may need up to $2\times$ the feature dimensionality of an unconstrained one to reproduce the same neighbourhood partition, so the fixed $N^*=3$ disadvantages the monotone models. 
	
	\paragraph{Sensitivity to CFO suppression}
	Because CFO is a strong but temperature- and time-varying fingerprint~\cite{xie2025robust}, we evaluate sensitivity to its blind suppression on the synthetic UED benchmark and both WiSig subsets. For each burst, we estimate
	\[
	\hat\nu=\frac{1}{2\pi}\overline{\arg\!\left(x[n]x^*[n-1]\right)}
	\]
	and apply $x[n]\leftarrow x[n]e^{-j2\pi\hat\nu n}$ before training and evaluation, leaving all other UED settings unchanged. The AUC$_{\mathrm{comp}}$ columns of Table~\ref{tab:ued} report the result. Compensation reduces AUC for every architecture. 
	
	On the synthetic task all fall to $50$--$53\%$, i.e., near chance, since CFO is essentially the only strong device-specific factor there. On WiSig the reductions are far smaller (about $1$--$14$ points) and all models remain well above chance, indicating that real emitters carry substantial fingerprint information beyond CFO.
	\section{Conclusion}
	\label{sec:conclusion}
	Polar MKAN imposes a simple structural constraint on RF-fingerprinting features, in which one latent coordinate depends only on magnitudes and two depend only on principal phases, with componentwise nondecreasing mappings in each connected block. The synthetic gain/CFO benchmark shows substantially stronger feature--factor alignment than the evaluated unpartitioned baselines, while WiSig shows a measurable UED-accuracy cost. Future work should separate partitioning from monotonicity experimentally, use circular or reference-invariant phase representations to remove the branch-cut limitation, and assess end-to-end attribution on flagged emitters.
	\bibliographystyle{IEEEtran}
	\bibliography{references}
\end{document}